\documentclass[conference]{IEEEtran}
\IEEEoverridecommandlockouts
\usepackage{cite}
\usepackage{amsmath,amssymb,amsfonts}
\usepackage{graphicx}
\usepackage{textcomp}
\usepackage{xcolor}
\usepackage{algorithm}
\usepackage{algpseudocode}
\usepackage{multirow}
\def\BibTeX{{\rm B\kern-.05em{\sc i\kern-.025em b}\kern-.08em
    T\kern-.1667em\lower.7ex\hbox{E}\kern-.125emX}}
\begin{document}

\title{Satellite Computing: A Case Study of \\ Cloud-Native Satellites
}

\author{Chao Wang\IEEEauthorrefmark{1}, Yiran Zhang\IEEEauthorrefmark{1}, Qing Li\IEEEauthorrefmark{2}, Ao Zhou\IEEEauthorrefmark{1}, and Shangguang Wang\IEEEauthorrefmark{1} \\

\IEEEauthorblockA{\IEEEauthorrefmark{1} \textit{State Key Laboratory of Networking and Switching Technology, Beijing University of Posts and Telecommunications}, \\ Beijing, 100876, China}
\IEEEauthorblockA{\IEEEauthorrefmark{2} \textit{School of Computer Science, Peking University}, Beijing, 100871, China}
c-wang@bupt.edu.cn, yiranzhang@bupt.edu.cn, liqingpostdoc@pku.edu.cn, \\ aozhou@bupt.edu.cn, sgwang@bupt.edu.cn \\
\textit{(Invited Paper)}

\thanks{This work was supported in part by NSFC (62032003 and U21B2016), and China Postdoctoral Science Foundation (8206300713).}}

\maketitle

\begin{abstract}
The on-orbit processing of massive satellite-native data relies on powerful computing power. Satellite computing has started to gain attention, with researchers proposing various algorithms, applications, and simulation testbeds. Unfortunately, a practical platform for deploying satellite computing is currently lacking. As a result, the industry needs to make relentless efforts to achieve this goal. We suggest using cloud-native technology to enhance the computing power of LEO satellites. The first main satellite of the Tiansuan constellation, BUPT-1, is a significant example of a cloud-native satellite. Prior to delving into the details of BUPT-1, we define the essential concepts of cloud-native satellites, i.e., the cloud-native load and cloud-native platform. Afterwards, we present the design scheme of cloud-native satellites, including the architecture of BUPT-1 and the experimental subjects it can support. Two validation tests are shown to reflect the operation and capability of BUPT-1. Besides, we predict possible research fields that could shape the future of satellites in the next decade.
\end{abstract}

\begin{IEEEkeywords}
Satellite Computing, Cloud-Native Satellite, BUPT-1, Cloud-Native Load, Cloud-Native Platform
\end{IEEEkeywords}

\section{Introduction}\label{part1}

As satellite performance continues to improve, the scope of satellite applications continues to expand \cite{r17}. The wide availability of satellite services has resulted in a significant surge in the amount of data stored onboard. The amount of data generated by remote sensing satellites worldwide far exceeds the capacity for timely downloading at present \cite{r24}. Even with the use of a higher frequency Ku/Ka band for downloading, it is impossible to complete the download of all data within a single satellite transit window. Moreover, the traditional interpretation method is no longer suitable for the characteristics of satellite images such as multi-sensor and multi-resolution. A solution has been proposed that combines image data processing with AI \cite{r25}. However, the execution of onboard AI requires the satellite to have sufficient computing power. Thus, there is a pressing need for satellite computing \cite{r22} to enable on-orbit processing of space-native data and to support emerging applications. Here, satellite computing is a broad concept, which includes computing, network, communication, control.

Industry and academia have initiated preliminary research on satellite computing in response to the pressing demand. Major organizations and companies around the world have proposed or conducted normative standards \cite{r16,r18,r19} and prior experiments for low earth orbit (LEO) constellations to establish the foundation for satellite computing \cite{r13,r14}. SpaceX has sent over 30,000 computers equipped with the Linux operating system into space as part of their satellite constellation. These computers have been stripped down and equipped with custom patches and drivers that interact with the hardware. To a computer, the Starlink satellites resemble a massive cluster of servers operating in space.

Scholars have also proposed various paradigms, frameworks, testbeds, and optimization algorithms for satellite computing. Bhattacherjee \textit{et al.} \cite{r2} initiated a scholarly discourse on the concept of satellite edge computing, exploring potential applications that could leverage this paradigm as well as identifying factors that could hinder its implementation. Denby \textit{et al.} \cite{r3} proposed using the nanosatellite orbital edge computing to tackle limitations of the ``bent-pipe'' architecture. Driven by remote sensing services, the authors presented a working mode of computational nanosatellite pipeline to eliminate the problem of single satellite energy constraints. Furthermore, the first orbital edge computational simulator is realized for mission simulation and autonomous control of satellites. Bhosale \textit{et al.} \cite{r4} designed a loosely coupled orchestration tool for joint path prediction, satellite selection, and task hand-off. Pfandzelter \textit{et al.} \cite{r5} analyzed the applicability of different application organization paradigms of the virtual machine, container, and serverless function according to the characteristics and requirements of satellite edge computing. Based on the previous work, they put forth the design principle for satellite simulation tools and designed a satellite edge computing testbed based on microVM \cite{r6}. Besides, the existing work mainly focuses on resource allocation \cite{r7}, computing offloading \cite{r8,r9}, service placement \cite{r10}, etc., which contribute to the improvement of satellite computing ecology from the theoretical perspective.

It is exciting to note is that in the past couple of years, satellites have made tremendous advancements in their performance and computing capabilities. For example, the LEO broadband communication satellite launched by China in January 2020 had a CPU clocked at 600MHz, a memory of 4GB, and a bus bandwidth of 1.28Gbps. In March 2022, China launched its first mass-developed LEO broadband communication satellite with an upgraded single-satellite CPU clocked at 1.2GHz, 8GB of memory, and a bus bandwidth of 2.5Gbps. These advancements demonstrate that satellites have begun computing and are providing new opportunities for the development of the entire satellite information industry. It is undeniable that correlational research on satellite computing is still in its infancy. In addition to the high technical requirements for satellite manufacturing, there is also a need to discuss the form in which computing tasks will be executed in space. At present, satellite computing is not readily accessible to the general public as a full-fledged commercial service. Besides, for the academic community, there is a pressing need for a genuine and open satellite computing platform to deploy and verify their frameworks, mechanisms, or algorithms. Satellite computing still has a long way to go from real industrial applications.

Together with academia and industry, we launched the Tiansuan constellation to build an open on-orbit computing platform \footnote{http://www.tiansuan.org.cn/index.html}. The first main satellite in the constellation, called BUPT-1, is of great significance as it provides a genuine practical platform for satellite computing, not just a theoretical or simulated one \cite{r1}. To give BUPT-1 the necessary computing power, we put a bunch of computing components on board. We package our applications as containers or services and upload them to those computing components. It is worth noting that cloud-native technology is a key factor in making all this work \cite{r23}, which is also the fundamental concept behind the design of BUPT-1. It lets us take advantage of resource virtualization, function generalization, and flexible, easy orchestration of BUPT-1. The successful operation of BUPT-1 demonstrates we have got a solid foundation for building an on-orbit open-source test platform for satellite computing.

We are convinced that satellites have an untapped potential that goes way beyond what we have seen so far. After large-scale LEO satellites and computational satellites, where is the future of satellites? What fields are going to rely on satellites in the years to come? And what kind of obstacles will it have to overcome along the way? This paper takes a deep dive into all of these questions.

The remainder of this paper is organized as follows. In Section \ref{part2}, we introduce the background of the Tiansuan constellation and satellite computing. Afterwards, we introduce our efforts to advance the satellite industry. The design concept and specific functions of BUPT-1 are elaborated in detail in Section \ref{part3}. Subsequently, we focus on key research fields for future satellites. Finally, the entire paper is summarized.

\section{Background}\label{part2}

\subsection{Tiansuan Constellation}

\begin{figure}
\centering
\includegraphics[width=1\columnwidth]{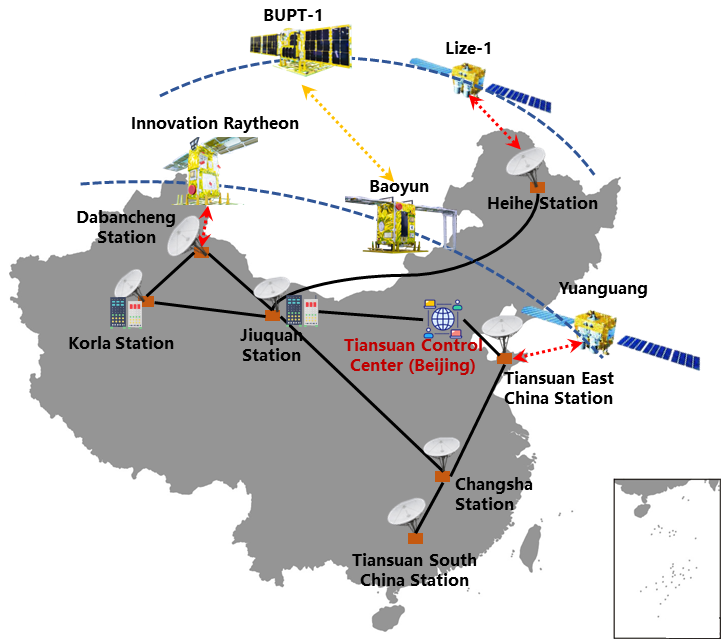}
\caption{The Architecture of Tiansuan Constellation}
\label{fig_ts}
\end{figure}

It is believed that satellites are the first stop for human beings to expand the living space of the earth. However, due to the high technical threshold of satellite research and development, as well as the high costs and risks associated with satellite operation and management, many scientific researchers find it difficult to get involved in the satellite industry. In October 2021, we initiated the Tiansuan Constellation Project, which is comprised of three phases with the objective of launching 6, 24, and 300 satellites respectively. It is currently in the first stage of the Tiansuan constellation. As shown in Fig. \ref{fig_ts}, there are already 5 satellites in orbit and operating normally, including a main satellite (BUPT-1), two auxiliary satellites (Baoyun and Innovation Raytheon), and two marginal satellites (Lize-1 and Yuanguang). Each satellite is equipped with the ability to communicate with ground stations. Moreover, to enhance the performance of satellite networks, efforts are underway to construct inter-satellite links and cloud-native satellite ground stations. The first phase is expected to be completed in 2023. In the first phase of the Tiansuan Constellation, the main tasks include but are not limited to developing the satellite-ground network, establishing the 6G core network, implementing space-air service computing, enabling cloud-native satellite edge computing, developing the satellite operating system, building the onboard AI acceleration platform, conducting device and load testing, integrating measurement, control and operation, and creating open platforms for public service capabilities.

The Tiansuan Constellation is dedicated to exploring new forms of satellite-ground computing, enabling satellite networks with satellite-ground computing, and connecting communication, navigation, remote sensing, and other space application needs with computing services. The goal of Tiansuan Constellation is to overcome innovation barriers and achieve heterogeneous constellation interconnection, satellite-ground integrated network, device load innovation, application data sharing, and other research objectives. Looking beyond, Tiansuan Constellation is expected to provide technical support for the discovery and exploration of human interstellar civilization and the development of interstellar networks. It serves as a platform for exchanging ideas and sharing resources for units or organizations interested in the space computing industry, as well as academia, research, and applications.

\subsection{Cloud-Native for Satellite Computing}

The widespread adoption of cloud computing and virtualization technology has driven the development of containerization technology, which has given rise to the emergence of cloud-native. Google's release of Kubernetes, an open-source container orchestration and management system, has further accelerated the adoption of cloud-native in commercial applications \cite{r15}. However, the application of cloud-native technology to satellites is still a relatively new concept and an initial attempt.

Cloud-native, with fundamental characteristics of microservices, containers, development\&operation\&maintenance (DevOps), and continuous delivery, endows satellites with more possibilities. Firstly, onboard microservices can be deployed, managed, and restarted independently, with the container serving as the carrier of microservice functions. Secondly, Docker is the most extensively used container engine, usually supervised and orchestrated by Kubernetes, which significantly improves the load balance between containers. Various container-based applications communicate through RESTful API. Then, DevOps provides the continuous delivery capability for cloud-native, strengthening the cooperation between developers and operators and facilitating the rapid deployment and production of applications. At last, continuous delivery ensures rapid delivery and feedback of cloud-native services through timely development and non-stop updating, reducing application release risks while ensuring service quality.

Using cloud-native technology for satellite computing offers unique advantages. Firstly, the cloud-native architecture is equipped with automation and integration functions that allow for the automatic deployment of user tasks on the satellite platform without the need for manual intervention or repeated execution. Secondly, cloud-native technology, which is based on micro-service architecture, can package computing tasks and requirements into micro-services, which are independent programs that can be released separately. This micro-service deployment method allows for greater flexibility and efficiency in managing the satellite computing platform. Furthermore, cloud-native can enhance the scalability and reliability of satellite computing. Containerized software running on a satellite platform operates independently without needing to consider other components. The isolation of the container ensures that the program runs smoothly without interference from external systems. Lastly, Cloud-native allows for satellite computing applications to run in containers, eliminating the need to consider differences in the underlying hardware of satellites. Containerized applications are also easy to migrate, simplifying the challenges associated with development, operation, and maintenance.

\section{BUPT-1 Satellite Design}\label{part3}

\subsection{Design Concept}

\begin{table*}[t]
\caption{Traditional Satellites versus Cloud-Native Satellites}
\centering
\begin{tabular}{|l|l|l|}
\hline
 & Traditional Satellite & Cloud-Native Satellite \\
\hline
\multirow{2}{*}{Deployment} & Debug the simulated environment in field & Packaged and ready-to-use container images \\ \cline{2-3}
                            & Communication protocol adapted to satellite-ground links & Unaware IP communication \\
\hline
\multirow{2}{*}{Scalability} & Unchangeable custom tasks & Update the image to switch applications \\ \cline{2-3}
                             & Terminal devices are fixed and cannot be dynamically managed & Pluggable device management framework \\
\hline
\multirow{2}{*}{Cost} & Telemetry and remote control managed by TT\&C station & Cloud unified management edge satellites \\ \cline{2-3}
                      & Downlink data managed by data transmission station & Load balancing and efficient use of constellation computing resources \\
\hline
\end{tabular}
\label{tab1}
\end{table*}

Cloud-native is a crucial concept in the design of Tiansuan constellation satellites, which significantly facilitates the development and management of onboard services, resources, and even satellite equipment. Cloud-native satellites have obvious advantages in application deployment, scalability, and cost-effectiveness compared to traditional satellites, as illustrated in TABLE \ref{tab1}. In a sense, cloud-native represents the future development direction of cloud computing and edge computing. As mentioned earlier, satellite computing encompasses computing, networking, and communication, while the cloud-native approach promotes the seamless integration and harmonization of these different components. In the practice of cloud-native satellites, cloud-native load, and cloud-native platform are two pivotal links:

\textbf{Cloud-Native Load:} The cloud-native load refers to the installation of a cloud-native platform that allows for the customization of payload functions using technologies such as Docker. It involves packaging the load functions into services using cloud-native technology, which are then bundled together with their necessary dependencies in containers. In simpler terms, it means that load capacity is not dependent on specific equipment for specific functions. Users can customize load functions as software definitions and deploy them as services and containers. By using computing hardware as the carrier and cloud-native technologies like microservices and Docker, cloud-native load provides users with local infrastructure alternatives that are similar to cloud services. This is different from the load function of traditional satellites which is mission-oriented. The load function is fixed during satellite development, so it is impossible to adjust the load capacity to meet user needs once the satellite is in space. Cloud-native load, on the other hand, offers the advantages of customizable functions, flexible services, more convenient management, and better cost performance.

\textbf{Cloud-Native Platform:} Cloud-native platform is a softening transformation that utilizes cloud-native technology to make satellite platform equipment, such as thermal control equipment, energy equipment, and attitude control equipment, more universal. The goal is to make satellite platform functions more compatible with each other. Different satellite manufacturers often have their own design standards and manufacturing engineering, which can lead to incompatibilities between the platform equipment and payloads. Various satellite platform equipment is connected to the onboard computer via a bus. In order for the payload device to work with the satellite hardware, it must be able to adapt to the interface protocol. Cloud-native platform solves this problem by creating a unified management platform for payloads and using software to define satellite platform equipment. It means that the differences between interface protocols can be ignored when the payload is deployed on the satellite platform. Instead, the payload only needs to make a simple call interface through the code. The realization of cloud-native platform greatly enhances the management and control capabilities of satellites, reduces the difficulty of application development and deployment, and benefits users, service providers, and satellite manufacturers.

\subsection{BUPT-1}

BUPT-1 was successfully launched and placed into orbit on January 15, 2023. It serves as the core node of the open platform for on-orbit space computing trials, actively practicing the concept of cloud-native. At present, it has successfully realized the cloud-native deployment of loads and is also working towards the direction of cloud-native platform. We present the detailed design of the cloud-native satellite architecture to intuitively understand BUPT-1. Moreover, cloud-native technology gives satellite computing platforms the ability to sustain extensive experiments.

\subsubsection{Cloud-Native Satellite Architecture}

\begin{figure*}
\centering
\includegraphics[width=0.85\textwidth]{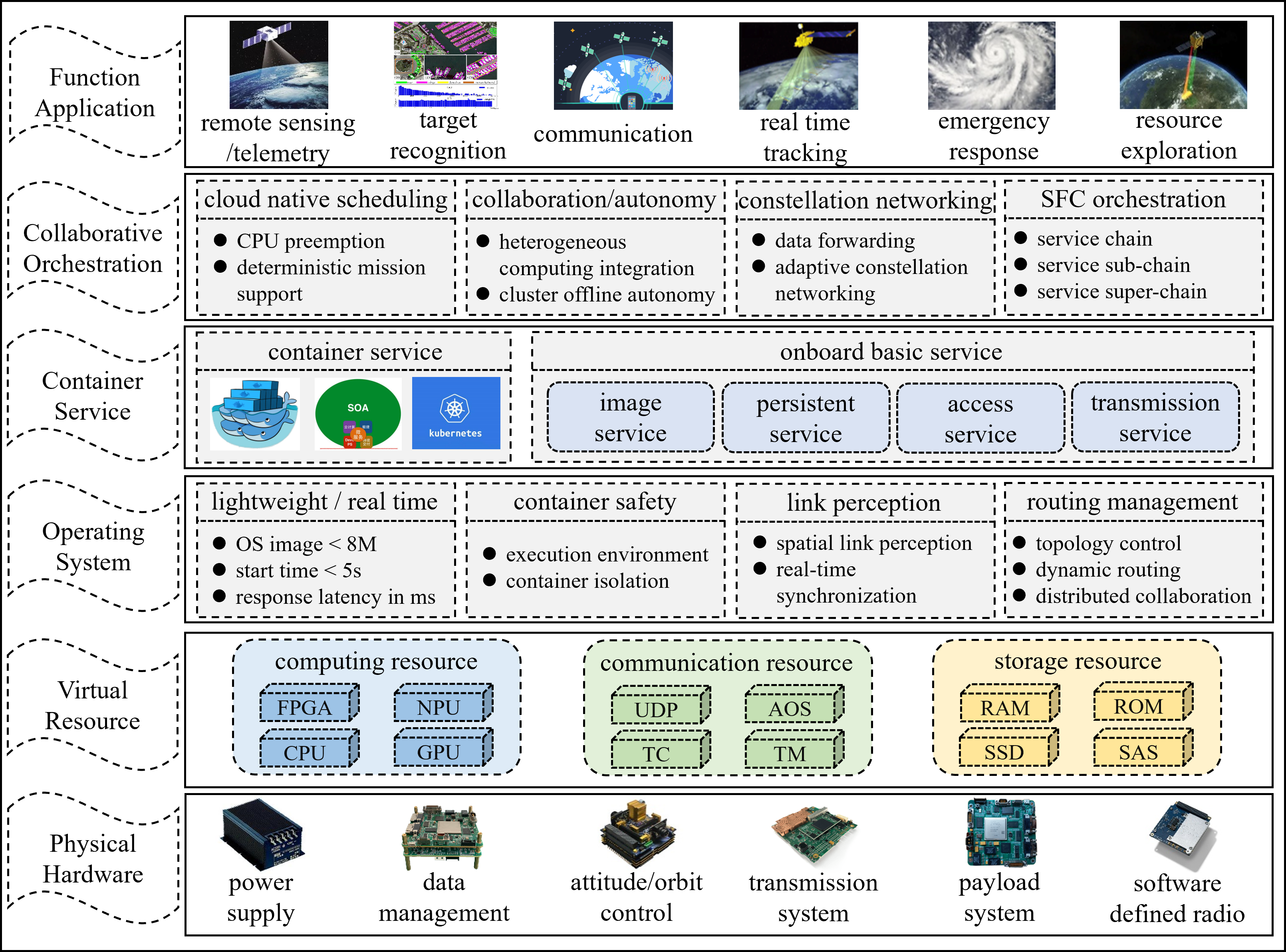}
\caption{Cloud-Native Satellite Architecture}
\label{fig1}
\end{figure*}

The cloud-native satellite architecture is shown in Fig. \ref{fig1}, which incorporates 6 function layers. Each layer showcases the components or capabilities that the cloud-native satellite should possess. To organically integrate the functions of all layers of the cloud-native satellite architecture, it provides a bottom-up vertical capability that includes infrastructure interconnection, integration of external resources, distributed system management, cross-domain service collaboration, interface capability openness, and application policy control. Furthermore, these capabilities also reflect the predominant functions of the corresponding layer. The cloud-native load is primarily concerned with managing and executing the functions of satellite payloads, which means that operating systems, container services, system orchestration, and certain virtual resources can all be implemented through the use of cloud-native load. Given that the cloud-native platform is designed for the entire satellite platform device, it supports the full range of functions associated with the cloud-native satellite architecture. It is worth noting that our focus in the next phase will be on the cloud-native platform.

\textbf{Physical hardware layer:} This layer contains the largest proportion of the essential platform equipment for the normal operation of satellites. For example, the power supply system is a component that produces, stores, adjusts, and allocates electricity to other systems of the satellite. Besides, the data management system, also known as the onboard computer, could be regarded as the brain of the satellite. It manages other onboard equipment through the satellite bus. General functions such as remote control, telemetry, payload management, and time reference are realized by the data management system. It is noteworthy that the software defined radio system is a sharp manifestation of the cloud-native platform, which is based on the wireless communication protocol defined by the software rather than hard wiring. The band width, air interface protocol, and function can be flexibly updated by software.

\textbf{Virtual resource layer:} Virtualization technology abstracts physical computing resources, communication resources, and storage resources into logical resource pools. The data management system then distributes these resources for applications. On-demand allocation of virtualized resources offers higher flexibility, preventing idle resources and resource fragmentation.

\textbf{Operating system layer:} Deploying an operating system for remote and dynamic satellites is non-trivial. It is necessary to develop adaptable operating system functions to accommodate the special environment of satellites and their characteristics, including but not limited to strong real-time/lightweight systems, container safety, spatial link perception, and routing management. The subsequent development and orchestration of applications can be carried out on the operating system.

\textbf{Container service layer:} Containers and microservices are the key technologies to realize cloud-native satellites, while Kubernetes and even Kubeedge are valid tools to manage containers and microservices. Applications and functions can be deployed on satellites in the form of containers or microservices for flexible development and orchestration. In addition to providing application container services, this layer also supports basic services such as image services, access services, transmission services, and persistent services.

\textbf{Collaborative orchestration layer:} This layer focuses on the collaborative orchestration and quality assurance of applications and services. Through an open capability interface, it provides orchestration services for cloud-native scheduling, constellation coordination and autonomy, cluster constellation networking, and service function chain to support user business needs effectively.

\textbf{Function application layer:} This layer covers a large number of businesses that LEO satellites and even satellite Internet are able to serve. Cloud-native satellites can provide customized and flexible services for specific applications through effective scheduling of onboard resources and coordination of various function layers. In the future, it should believe that with the improvement of cloud-native satellite capability, the business scope will be more extensive.

\subsubsection{Potential Experimental Exploration}

Based on the containerized deployment, BUPT-1 supports extensive experiments. We introduce several experiments being carried out by BUPT-1. It does not mean that we want to limit the scope of experiments supported by BUPT-1.

\textbf{Computing:} Satellite computing has been proposed for some time. Satellite computing, as a supplement to the capacity of ground hotspots, can relieve the load pressure on ground servers without worrying that satellite infrastructure would be damaged by natural disasters \cite{r27}. It has intuitive significance for processing satellite native data, especially the information processing of remote sensing images will be more accurate. As mentioned above, work on satellite computing has been conducted from the perspectives of satellite computing systems, simulation tools, and resource and service optimization. However, the potential of satellite computing goes beyond that, especially with BUPT-1 providing the possibility of on-orbit computing for emerging businesses. Firstly, AI in space: Remote sensing is the mainstream business of the majority of current LEO satellites, generating nearly 100TB of image data within the global daily. While it is common to download inference tasks to ground stations, this creates immense bandwidth pressure on satellite ground links. Additionally, overcast or cloudy weather can cause remote sensing satellites to generate invalid images, occupying limited downlink bandwidth. Onboard AI can infer remote sensing images on the spot, or select valuable images to send to the ground station for further processing. Secondly, federated learning in space: Most of the satellite data has to be downlinked to the ground station for further processing, which faces many challenges. On the one hand, the satellite-ground link is fragile, and data transmission faces high latency and interruption. On the other hand, data transmission in the satellite-ground link is vulnerable to leakage and eavesdropping. As a distributed machine learning mechanism, federated learning can directly process the information collected by satellites and protect the data privacy of satellites. Therefore, federated learning will have broad prospects in the processing acceleration and security guarantee of satellite computing.

\textbf{Networking:} Satellite networking is a mandatory requirement for 6G \cite{r26}. With an increase in satellite launches, the Tiansuan constellation is also moving towards this objective, while BUPT-1 supports some network-related experiments effectively, including traffic routing at the network layer, congestion control at the transport layer, and even remote sensing and target recognition at the application layer, as well as the core network. We believe that the satellite network will significantly promote the construction and improvement of 6G, in which the deployment and optimization of the onboard core network is the key link. In the early stages, we proposed a distributed cloud-edge core network architecture for 6G requirements \cite{r11}. The cloud core network is deployed on the ground and acts as a central coordination node, while the edge core network is deployed on each satellite and has the capability of self-management. In addition, we have confirmed the feasibility of deploying the core network on satellites \cite{r12}. The onboard core network will reduce the signaling interaction latency of the control plane, realize real-time services, and avoid signaling storms. With the rapid construction of space infrastructure, traditional constellation management methods have become obsolete. Utilizing the onboard core network to directly manage the network, computing, and storage resources will greatly improve the operating efficiency of the constellation. We will also continue to carry out a series of onboard core network experiments on BUPT-1 to verify its service capabilities.

\textbf{Measurement:} Access to real satellite data sets, including satellite-native data such as energy, heat consumption, and temperature, as well as application data like remote sensing images and monitoring data, is unusual for the academic community. This has posed a bottleneck for many researchers when conducting related experiments. Tiansuan Constellation is an open computing platform jointly built by multiple parties. With the improvement of relevant regulations and the maturity of technology, people who have a demand for satellite data sets will have more opportunities to access them. Currently, a large number of interesting measurement experiments have been carried out on BUPT-1, including satellite-ground link transmission (delay, bandwidth, jitter, packet loss rate, etc.), energy collection, storage and release, the relationship between energy consumption and load, and different load tradeoffs, etc. Since the Tiansuan constellation is still in the initial stage, the data collected by measurement experiments currently only comes from the single satellite of BUPT-1. It is believed that as the cloud-native satellite computing platform matures, Tiansuan Constellation will continue to provide rich and valuable satellite data sets.

\subsection{Tests}

\subsubsection{Brain-Satellite Interaction}

\begin{figure}
\centering
\includegraphics[width=1\columnwidth]{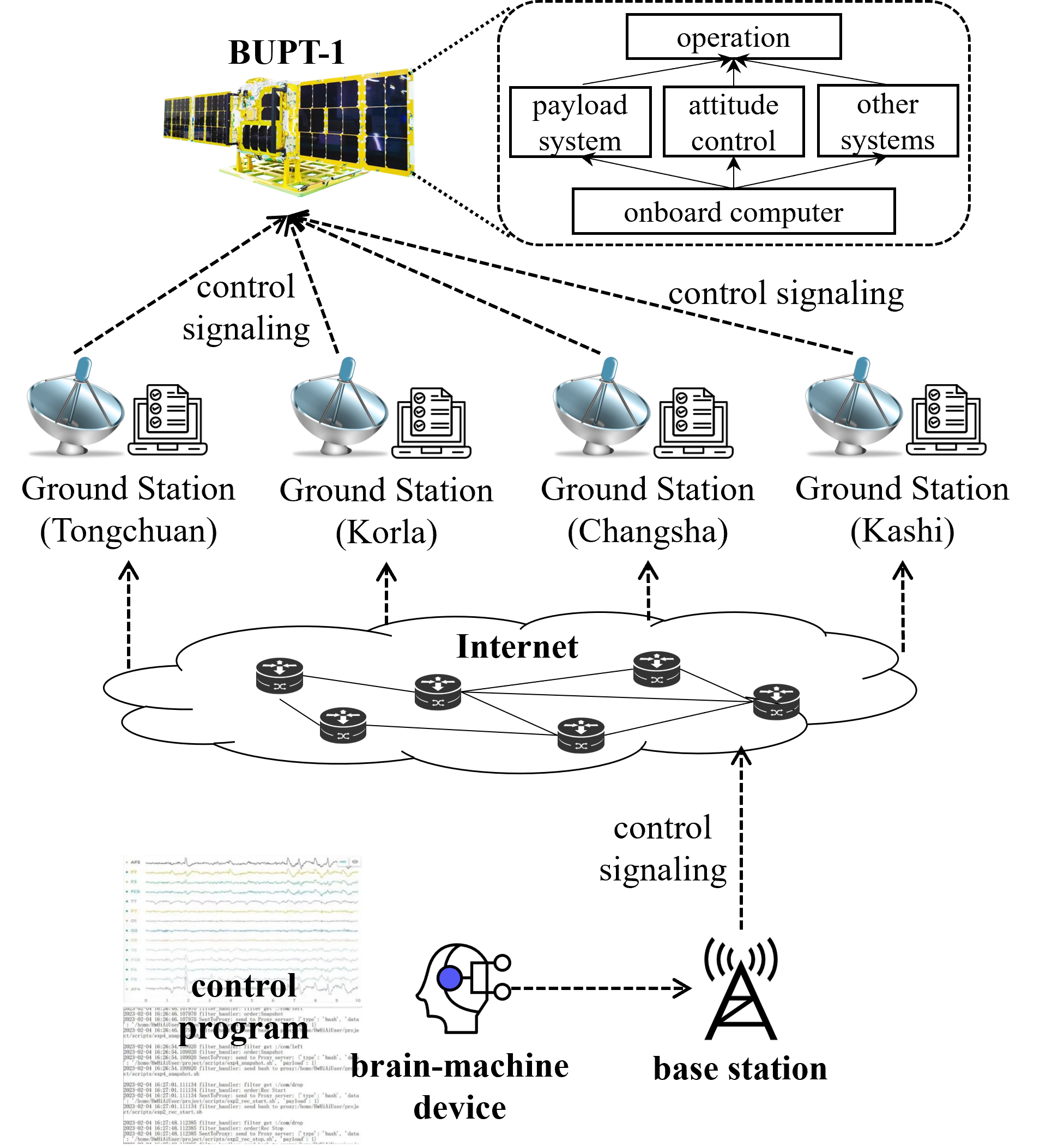}
\caption{Brain-machine device controls the BUPT-1 satellite.}
\label{fig2}
\end{figure}

From January 30 to February 4, 2023, we carried out several consecutive tests on the control of the BUPT-1 satellite using a brain-machine device. Experimenters wore the device and control BUPT-1 using brain signals through the Internet. The experimental structure/process is shown in Fig. \ref{fig2}. During the test, control commands were successfully executed using the brain-machine device. The cloud-native load also enables functional testing, including onboard payload application control, video capture and transmission, and satellite measurement and control. The test utilized a pre-trained AI model to extract cognitive information from EEG signals and completed the mapping of cognitive information to satellite services based on reinforcement learning algorithms. The mapping results were transmitted to BUPT-1 by calling the service interface. The ground station then provided services according to the content, thus completing the conversion of satellite measurement and control commands. Finally, the ground station encapsulated the satellite application control primitives into uplink remote control commands and sent them to BUPT-1 for interpretation and execution of the commands.

The success of the human brain-controlled satellite test demonstrates that the BUPT-1 has the capability to access and control satellites through the brain-machine interface, showcasing the flexibility and scalability of the cloud-native technology system adopted by BUPT-1 when applied to satellite measurement and control. This achievement provides a new potential technical reference for future satellite measurement and control and management modes.

\subsubsection{Real-time Video Transmission}

\begin{figure}
\centering
\includegraphics[width=1\columnwidth]{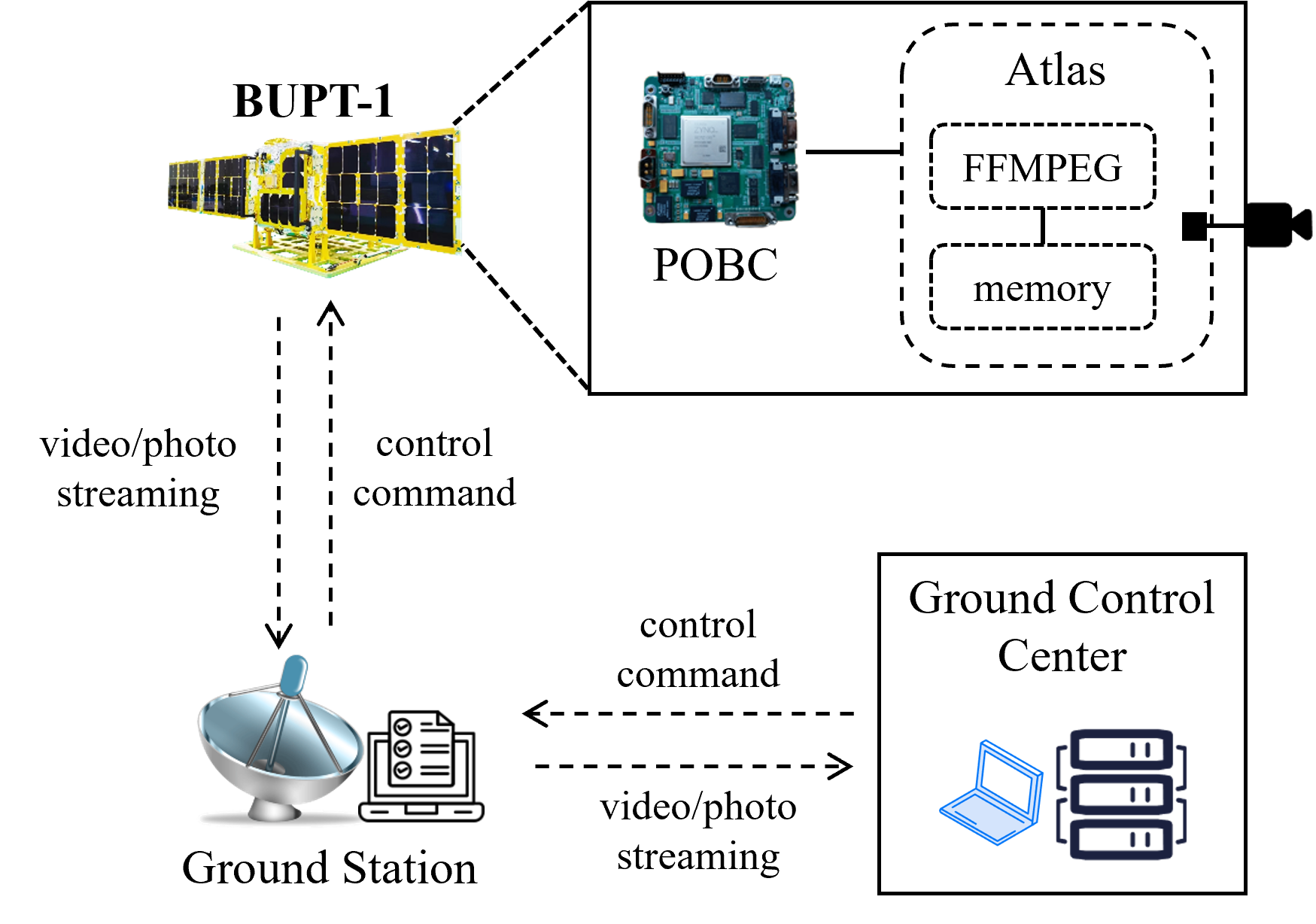}
\caption{Satellite-ground real-time video transmission.}
\label{fig3}
\end{figure}

BUPT-1 is equipped with four primary payloads primarily for testing, with the primary payload Atlas being deployed for real-time video transmission experiments, as shown in Fig. \ref{fig3}. The test includes sub-tests such as video recording and transmission, live streaming, and photography. The video recording and transmission test can be controlled to start and end through remote commands or can be scheduled to end after a fixed period of time. Video files can be saved in Atlas memory or can be transmitted to the ground after modulation using FFMPEG. The real-time live streaming test requires the satellite to establish a connection with the ground station and remote commands to notify the payload onboard computer (POBC) to charge. The Atlas camera records video and transmits data packets to POBC. After receiving the payload's real-time downlink message, POBC sends a response message. Atlas waits for confirmation from POBC before sending the next message while the video stream is transmitted in real-time to the ground station, achieving a live-streaming effect. The photography test controls the Atlas camera to take photos as needed through remote commands. Photos can be stored in memory for onboard processing or transmitted directly to the ground station. The onboard processing of photos relies on satellite inference capabilities, which are implemented through the cloud-native load. The success of the real-time video transmission test relies on both the processing ability of the payload and the satellite's ability to perform in-orbit calculations using images and video data. It also depends on the high-quality transmission capability of the satellite-ground links.

\section{Open Problems}\label{part4}

In this section, we will take a look at the future of satellite development, with a focus on the latest research areas and a rational analysis of open problems, hoping to provide effective references for researchers.

\subsection{Satellite Infrastructure}

In order for a satellite to enter its predetermined orbit, it must first be carried by a launch rocket. The frenzied construction of large-scale LEO constellations has been made possible by the reduction of satellite manufacturing costs and the maturation of rocket reuse technology. The scale of Starlink, which currently has 4,000+ satellites in orbit, is truly impressive and is achieved through the effective use of rocket reuse and the ``one rocket with multiple satellites'' approach. The Falcon 9 rocket has successfully completed 15 satellite delivery missions to date, and SpaceX has also achieved more than 100 rocket recoveries. China also has relatively mature rocket reuse technology but is pursuing an even higher goal: the complete reuse of rocket power systems. Traditional rocket recovery only allows for the reuse of part of the power system, while complete reuse of the power system is the ultimate goal. Therefore, further breakthroughs in rocket recovery technology will be a pressing issue for manufacturers and researchers alike.

As the amount of space-native data continues to explode, the quality of data downlink becomes a bottleneck for satellite-ground links and ground stations. Limited uplink and downlink bandwidth is a strict constraint for satellite-ground links, particularly for civil or academic satellites, where available bandwidth is highly asymmetrical. Efficiently downloading data with limited capacity and unstable links is a significant challenge. Taking the Tiansuan constellation as an example, where the communication between the in-orbit satellites and the ground station primarily utilizes the X-band. Currently, the uplink rate in this frequency band is limited to 0.1 Mbps-1 Mbps, while the downlink rate is about 100 Mbps - 600 Mbps. As a result, there is a significant asymmetry between the uplink and downlink speeds. To eliminate the transmission bottleneck in the satellite-ground link, the Tiansuan Constellation plans to apply for the utilization of Ku/Ka bands in the future. This would enable the uplink rate to exceed 200 Mbps, ensuring a more balanced and improved communication performance.

Ground stations, on the other hand, serve as the medium for communication between users and satellites, but they face challenges such as high construction costs, multi-satellite competition, limited communication time, and susceptibility to geological disasters. Existing research \cite{r20,r21} proposes distributed ground stations, ground station function design, adaptive downlink scheduling, and small-scale ground stations based on commercial hardware to optimize satellite-ground communication quality and ground station service capabilities. Moreover, the ground station serves as the channel connecting satellites and ground networks, as well as the portal for us to communicate with satellites. The Tiansuan constellation requires a wide range of ground stations to allow users to access the service anytime and anywhere. Building cloud-native ground stations is a vital infrastructure for adapting to and serving cloud-native satellites. More investment and effort are required to enhance the utilization of limited spectrum resources and ground station performance.

\subsection{Satellite Edge Computing}

Satellite edge computing is a mobile edge computing application that is used in a specific scenario \cite{r28}. Satellite edge computing leverages the global mobility of satellites to bring computing services to all corners of the world, while satellites benefit from satellite edge computing to supplement their limited computing power. Despite the limited progress made in the design of the underlying satellite edge computing architecture and the optimization of upper service deployment and processes, the industry has made significant advancements. In particular, the BUPT-1 is a pioneer in satellite edge computing, and many verification tests have been successfully conducted.

There is no denying that satellite edge computing offers a multitude of intuitive and varied benefits. In computing offloading, ground users with limited computing resources can offload computing tasks to LEO satellites with satellite edge computing capabilities to reduce task processing latency and resource overhead. Even if ground users are in remote areas or if the ground infrastructure is damaged, satellite edge computing can rely on the inherent characteristics of LEO satellites to provide computing services and emergency communication capabilities for any user. In data caching/processing, LEO satellites, especially remote sensing satellites, require constant data processing, which puts immense pressure on satellite-ground links and ground stations. On the one hand, satellite edge computing caches data in satellites or ground stations to avoid the retransmission of similar content and reduce network traffic. On the other hand, satellite edge computing enables LEO satellites to have onboard computing capabilities, processing space-native data on the satellite, thereby reducing the amount of downloading. In security, compared to traditional cloud computing, satellite edge computing offloads user information to multiple edge LEO satellites to avoid centralized processing of user information. In summary, satellite edge computing will contribute to the future satellite ecology in terms of satellite-native data processing, resource optimization, etc.

\subsection{Onboard Core Network}

The satellite network is a crucial part of the space segment in the space-ground integrated networks (SGIN), and it needs to have communication capabilities that match those of terrestrial networks. With this in mind, the deployment of the core network on satellites has become a pressing issue. In 5G, the core network is designed and deployed based on a service-oriented architecture, but it may not be suitable for satellite deployment due to its centralized architecture for terrestrial communication, the static rigidity of 5G network slices, and limited business scenarios. While there is no official standard for the architecture and functions of the 6G core network yet, it should be developed to enhance communication and operate in space-air-ground-sea scenarios. The 6G core network functions can encapsulate satellite communication capabilities (spectrum, interfaces, protocols, etc.) to make communication service-oriented. The deployment of core networks on satellites presents opportunities for network element function simplification, process logic design, interface function simplification, and service combination supply, which are all necessary ways to improve satellite networks and SGIN capabilities. Therefore, deploying an onboard core network will add value to satellite communications and the construction of the SGIN.

\subsection{Onboard Operating System}

Satellite development is evolving towards ubiquitous, intelligent, and universal capabilities. The rapid advancement of satellite software has led to innovation in onboard hardware. The use of commercial hardware in satellites not only reduces manufacturing costs but also enhances flexibility in task deployment with the unified architecture of onboard computers and payloads. Consequently, research and development of onboard operating systems have become a popular topic. The development of a new generation of satellite operating systems should overcome the limitations of traditional onboard computer functions. While improving performance, it should seamlessly integrate with the existing Linux software ecosystem. Furthermore, the onboard operating system, based on commercial hardware, should provide a fault-tolerant mechanism at the software level to avoid interference from unreliable environments. Lastly, the new generation satellite operating system should include isolation mechanisms for a large number of onboard tasks to prevent interference and provide scheduling functions based on task priorities.

\subsection{Space-Ground Integrated Network}

As B5G and 6G technologies are introduced, the creation of the SGIN has become an inevitable trend \cite{r29}. Many major countries and companies have issued policies or plans to promote the construction of the SGIN. However, the SGIN is still in its early stages and faces numerous challenges. The emerging SGIN paradigm presents a plethora of research opportunities. Firstly, the management of the vast number of massive satellites in the space segment is a significant challenge. Most of the satellites used in building the SGIN are LEO satellites, which are highly dynamic and heterogeneous. The LEO constellation is expected to be equipped with inter-satellite links, which will further increase the system's complexity. Secondly, intelligent resource management and orchestration is a crucial topic for the SGIN. Global users have varied service quality requirements, and satellite networks have different resource capacities, communication qualities, and topology stability compared to terrestrial networks. Hence, there is a need for more reliable and efficient resource perception and orchestration algorithms to manage SGIN resources. Thirdly, the SGIN's dynamic heterogeneity and service stability pose a prominent challenge. To provide users with stable network services on demand, it is necessary to dynamically perceive the end-to-end network resources in the SGIN. The ever-changing satellite topology creates difficulties in matching complex services and resources, end-to-end routing, and coordinated transmission. Lastly, the security of the SGIN must be considered. Satellites, inter-satellite links, and satellite-ground links are all exposed in open space, making them vulnerable to cyber-attacks like eavesdropping and information tampering. However, there are no specific security standards for satellite networks at present.

\subsection{Others}

The development of 6G networks requires the integration of satellite communication, perception, and computing capabilities. In the past, the development of these three components has been independent of each other. However, 6G imposes more rigorous requirements on network latency, bandwidth, computing power, speed, connection density, and other indicators, making it necessary to combine the three components. Unfortunately, there is currently no unified standard in terms of system architecture, enabling technology, and evaluation indicators for this integration. It is essential to take into account the similarities and differences between communication, perception, and computing when integrating them. To achieve this, it may be necessary to rely on specific technologies such as deep learning, reinforcement learning, federated learning, and software-defined satellites.

Satellite development is a complex and interdisciplinary undertaking that requires expertise in various fields such as computer science, networking, communications, automation, electronics, and mechanics. Specialists with diverse academic and industrial backgrounds must focus on different areas, such as satellite attitude and orbit control systems, energy systems, thermal control systems, measurement and control systems, and operating systems to contribute to the development and enhancement of the satellite industry ecosystem.

\section{Conclusion}\label{part5}

Satellite computing is a key enabler for realizing a true space-air-ground-sea integrated network for 6G. We advocate that cloud-native is the prominent direction for the future of computing satellites and are actively exploring the key technologies of cloud-native satellites. This paper introduces BUPT-1, an open cloud-native satellite of the Tiansuan constellation. Subsequently, we present the design concept of Tiansuan satellites and elaborate on the key aspects of designing the cloud-native satellite. Afterward, we provide some interesting tests deployed on BUPT-1, which fully demonstrate its capabilities. Finally, we look at the future development prospects of satellites. We call on relevant researchers to pay attention to the application of cloud-native technology in satellites, and work together with us to build an open Tiansuan constellation platform.


\begin{thebibliography}{00}

\bibitem{r17} P. Zhang, Z. Pang and W. Jiang, ``Application of Communication, Navigation and Remote Sensing Integration in Water Safety Monitoring in Special Areas,'' {\em Proceedings of the International Conference on Geoinformatics}, pp. 1-3, 2022.

\bibitem{r24} B. Denby, K. Chintalapudi, R. Chandra, B. Lucia and S. A. Noghabi, ``Kodan: Addressing the Computational Bottleneck in Space,'' {\em Proceedings of the International Conference on Architectural Support for Programming Languages and Operating Systems}, pp. 392-403, 2023.

\bibitem{r25} C. Zheng, J. Wang, A. Ma and Y. Zhong, ``AutoLC: Search Lightweight and Top-Performing Architecture for Remote Sensing Image Land-Cover Classification,'' {\em Proceedings of the International Conference on Pattern Recognition}, pp. 324-330, 2022.

\bibitem{r22} Z. Zhang, W. Zhang and F. -H. Tseng, ``Satellite Mobile Edge Computing: Improving QoS of High-Speed Satellite-Terrestrial Networks Using Edge Computing Techniques,'' {\em IEEE Network}, vol. 33, no. 1, pp. 70-76, 2019.

\bibitem{r16} 3GPP, ``Study on New Radio (NR) to Support Non Terrestrial Networks: 38.811,'' {\em 3rd Generation Partnership Project (3GPP)}, 2017. https://www.3gpp.org/ftp/specs/archive/38\_series/38.811.

\bibitem{r18} 3GPP, ``Solutions for NR to Support Non-Terrestrial Networks (NTN): Non-Terrestrial Networks (NTN) Related RF and Co-existence Aspects: 38.863,'' {\em 3rd Generation Partnership Project (3GPP)}, 2018. https://www.3gpp.org/ftp/specs/archive/38\_series/38.863.

\bibitem{r19} 3GPP, ``Study on Management Aspects of IoT NTN Enhancements: 28.841,'' {\em 3rd Generation Partnership Project (3GPP)}, 2022. https://www.3gpp.org/ftp/Specs/archive/28\_series/28.841.

\bibitem{r13} Y. Guo, Q. Li, Y. Li, N. Zhang and S. Wang, ``Service Coordination in the Space-Air-Ground Integrated Network,'' {\em IEEE Network}, vol. 35, no. 5, pp. 168-173, Sept./Oct. 2021.

\bibitem{r14} C. Ravishankar, R. Gopal, N. BenAmmar, G. Zakaria and X. Huang, ``Next-Generation Global Satellite System with Mega-Constellations,'' {\em International Journal of Satellite Communications and Networking}, vol. 39, pp. 6-28, 2021.

\bibitem{r2} D. Bhattacherjee, S. Kassing, M. Licciardello and A. Singla, ``In-orbit Computing: An Outlandish thought Experiment?'' {\em Proceedings of the ACM Workshop on Hot Topics in Networks}, pp. 197-204, 2020.

\bibitem{r3} B. Denby and B. Lucia, ``Orbital Edge Computing: Nanosatellite Constellations as a New Class of Computer System,'' {\em Proceedings of the International Conference on Architectural Support for Programming Languages and Operating Systems}, pp. 939-954, 2020.

\bibitem{r4} V. Bhosale, K. Bhardwaj and A. Gavrilovska, ``Toward Loosely Coupled Orchestration for the LEO Satellite Edge,'' {\em Proceedings of the USENIX Workshop on Hot Topics in Edge Computing}, pp. 1-7, 2020.

\bibitem{r5} T. Pfandzelter, J. Hasenburg and D. Bermbach, ``Towards a Computing Platform for the LEO Edge,'' {\em Proceedings of the International Workshop on Edge Systems, Analytics and Networking}, pp. 43-48, 2021.

\bibitem{r6} T. Pfandzelter and D. Bermbach, ``Celestial: Virtual Software System Testbeds for the LEO Edge,'' {\em Proceedings of the International Middleware Conference}, pp. 69-81, 2022.

\bibitem{r7} X. Xu, H. Zhao, C. Liu, C. Fan, Z. Liang and S. Wang, ``On the Aggregated Resource Management for Satellite Edge Computing,'' {\em Proceedings of the IEEE International Conference on Communications}, pp. 1-6, 2021.

\bibitem{r8} L. Cheng, G. Feng, Y. Sun, M. Liu and S. Qin, ``Dynamic Computation Offloading in Satellite Edge Computing,'' {\em Proceedings of the IEEE International Conference on Communications}, pp. 4721-4726, 2022.

\bibitem{r9} H. Fang, Y. Jia, Y. Wang, Y. Zhao, Y. Gao and X. Yang, ``Matching Game based Task Offloading and  Resource Allocation Algorithm for Satellite Edge Computing Networks,'' {\em Proceedings of the International Symposium on Networks, Computers and Communications}, pp. 1-5, 2022.

\bibitem{r10} L. Cheng, F. Tang and X. Li, ``Mobility- and Load-Adaptive Controller Placement and Assignment in LEO Satellite Networks,'' {\em Proceedings of the IEEE Conference on Computer Communications}, pp. 1-10, 2021.

\bibitem{r1} S. Wang, Q. Li, M. Xu, X. Ma, A. Zhou and Q. Sun, ``Tiansuan Constellation: An Open Research Platform,'' {\em Proceedings of the IEEE International Conference on Edge Computing}, pp. 94-101, 2021.

\bibitem{r23} J. Larrea, M. K. Marina and J. E. v. d. Merwe, ``Nervion: a Cloud Vative RAN Emulator for Scalable and Flexible Mobile Core Evaluation,'' {\em Proceedings of the Annual International Conference on Mobile Computing and Networking}, pp. 736-748, 2021.

\bibitem{r15} R. Botez, C.-M. Iurian, I.-A. Ivanciu and V. Dobrota, ``Deploying a Dockerized Application With Kubernetes on Google Cloud Platform,'' {\em Proceedings of the International Conference on Communications}, pp. 471-476, 2020.

\bibitem{r11} Y. Li, J. Huang, Q. Sun, T. Sun and S. Wang, ``Cognitive Service Architecture for 6G Core Network,'' {\em IEEE Transactions on Industrial Informatics}, vol. 17, no. 10, pp. 7193-7203, Oct. 2021.

\bibitem{r12} R. Xing, X. Ma, A. Zhou, S. Dustdar and S. Wang, ``From earth to space: A first deployment of 5G core network on satellite,'' {\em China Communications}, pp. 1-11, 2022, doi: 10.23919/JCC.2023.00.001.

\bibitem{r20} D. Vasisht and R. Chandra, ``A Distributed and Hybrid Ground Station Network for Low Earth Orbit Satellites,'' {\em Proceedings of the the ACM Workshop on Hot Topics in Networks, Virtual Event}, pp. 190-196, 2020.

\bibitem{r21} D. Vasisht, J. Shenoy and R. Chandra, ``L2D2: low latency distributed downlink for LEO satellites,'' {\em Proceedings of the Conference on Applications, Technologies, Architectures, and Protocols for Computer Communication}, pp. 151-164, 2021.

\bibitem{r27} Q. Li, S. Wang, X. Ma, Q. Sun, H. Wang, S. Cao and F. Yang, ``Service Coverage for Satellite Edge Computing,'' {\em IEEE Internet of Things Journal}, vol. 9, no. 1, pp. 695-705, 2022.

\bibitem{r26} Y. C. Chou, X. Ma, F. Wang, S. Ma, S. H. Wong and J. Liu, ``Towards Sustainable Multi-Tier Space Networking for LEO Satellite Constellations,'' {\em Proceedins of the IEEE/ACM International Symposium on Quality of Service}, pp. 1-11, 2022.

\bibitem{r28} T. Kim, J. Kwak and J. P. Choi, ``Satellite Edge Computing Architecture and Network Slice Scheduling for IoT Support,'' {\em IEEE Internet of Things Journal}, vol. 9, no. 16, pp. 14938-14951, 2022.

\bibitem{r29} F. Lyu, W. Xu, Q. Yuan and K. Suto, ``Space-Air-Ground Integrated Networks for Future IoT: Architecture, Management, Service and Performance,'' {\em Peer-to-Peer Networking and Applications}, vol. 14, no. 5, pp. 3265-3267, 2021.

\end{thebibliography}
\end{document}